# The impact of ionic contribution to dielectric permittivity in 11CB liquid crystal and its colloids with BaTiO₃ nanoparticles

Joanna Łoś[1], Aleksandra Drozd-Rzoska[1], Sylwester J. Rzoska[1], Krzysztof Czupryński[2]

[1] Institute of High Pressure Physics Polish Academy of Sciences, X-PressMatter Lab, ul. Sokołowska 29/37, 01-142 Warsaw, Poland.

[2] Military University of Technology, Faculty of Advanced Technologies and Chemistry, ul. gen. Sylwestra Kaliskiego 2, 00-908 Warsaw, Poland

**corresponding author**: joalos@unipress.waw.pl

**ORCIDs**: JŁ  0000-0001-9102-7425;  ADR  0000-0001-8510-2388;  SJR  0000-0002-2736-2891; KC 0000-0002-1785-7786

**Abstract**

The report shows the temperature behavior of the real part of dielectric permittivity in the static (dielectric constant ) and low-frequency (LF) domains in bulk samples of 11CB and its BaTiO$_3$-based nanocolloids. The study covers the Isotropic Liquid (I), Nematic (N), Smectic A (SmA), and Solid Crystal (Cr) phases. For each phase, the dominance of pretransitional fluctuations, significantly moderated by nanoparticles, is shown. The authors consider separate focuses on the dielectric constant $\varepsilon(T)$ evolution in the static domain, yielding mainly response from permanent dipole moment and their arrangement, and in the low-frequency (LF) domain $\Delta\varepsilon'(f) = \varepsilon'(f) - \varepsilon$ (where $\varepsilon'(f)$ is for the real part of dielectric permittivity in the LF domain), which is associated solely with ionic-related polarization mechanisms. All of these led to new experimental evidence concerning I-N, N-SmA, and SmA-Solid transitions, focusing on the strength and extent of pretransitional effects, critical exponents, and phase transitions discontinuities. The strong evidence for pretransitional effects near the SmA – Cr transition is notable, particularly regarding $\Delta\varepsilon'(f, T)$. Studies are supplemented by the discussion of DC electric conductivity - a parameter also related to the LF domain. Finally, the validity of the relation $\varepsilon'(f) = Af^{-3/2} + \varepsilon$ (where $f$ stands for frequency, and A is a constant parameter), often used for discussing dielectric spectra in LC compound and its nanocolloids in the LF domain is examined.

# 1. Introduction

Liquid crystalline (LC) materials are essential for a myriad of applications, exploring their extraordinary sensitivity to external disturbances, especially the electric field [1-3]. LC compounds also constitute a unique experimental model system for fundamental research [1, 2, 4]. Regarding the latter, the isotropic liquid (I) phase for rod-like LC is considered as a model for studying the vitrification and glass transition phenomenon [4-10]. Notable are similarities between Smectic A (SmA) – Nematic (N) transition and those occurring in superconductors [11, 12]. The cognitive attention attracts the freezing of single symmetry elements in subsequent mesophases on cooling [1, 4].

Nanoparticles (NPs) are a unique type of solids. for which the reduction in grain size below 100 nm leads to the appearance of unique, even exotic, properties – strongly manifested in interactions with surrounding them matrix systems [13]. Liquid crystals are especially suitable for this role due to the mentioned extraordinary sensitivity to even slight disturbances. The research carried out for several decades has shown unusual properties in LC-based nanocolloids (low concentrations of NPs) and nanocomposites (higher concentrations of NPs), in fact leading to the emergence of a new research area in the field of liquid crystals and soft matter physics [14, 15]. Notwithstanding, the area of LC+NPs systems is still on the exploratory path beginning.

For LC or LC+NPs, the interaction with the external electric field is essential [1, 2], which indicates the primary role of broadband dielectric spectroscopy (BDS) [16] as the experimental monitoring method. Particularly important are studies of the dielectric constant ($\varepsilon$), directly related to the ability of molecules to interact with the electric field. Temperature evolution of the dielectric constant allows to discern the prevalence of different arrangements of permanent dipole moment coupled to molecules, namely: $d\varepsilon/dT < 0$ is related to the 'parallel' arrangement, i.e., dipole moment follows the electric field, and $d\varepsilon/dT > 0$ is for their antiparallel ordering [17]. Both in LC and LC+NPs systems, studies of dielectric constant and related properties revealed the dominant influence of multimolecular pretransitional fluctuations in the isotropic liquid and LC mesophases [5, 7, 18-21]. It is manifested via long-range pretransitional effects associated with the weakly-discontinuous character of $Isotropic \leftrightarrow Nematic$, $Isotropic \leftrightarrow Smectic$, $Nematic \leftrightarrow Smectic$, phase transitions. Such behavior is manifested via pretransitional effects explained within the *Physics of Critical Phenomena* [4]. For dielectric constant the following general relation describing such behavior can be concluded [18-24]:

$$\varepsilon(T) = \varepsilon^* + a|T - T^*| + A|T - T^*|^{1-\alpha} \tag{1}$$

where $\varepsilon^*$ and $T^*$ describe the extrapolated continuous phase transition located beyond the mesophase in which the given pretransitional effect is scanned; in the isotropic phase $T > T^C = T^* + \Delta T^*$, where $T^C$ is for the clearing temperature, i.e., the isotropic – LC mesophase transition temperature, $\Delta T^*$ is the metric of the phase transition discontinuity, $a$ and $A$ are constant parameters, and the exponent $\alpha$ is related to the specific heat pretransitional anomaly.

Eq. 1 was first introduced in refs. [18, 22], by the analogy to the behavior occurring in the homogeneous phase of critical binary mixtures of limited miscibility. The latter was theoretically derived by Oxtoby et al. [25], using the droplet model approximation for precritical fluctuations, and by Sengers et al. [26], considering the internal energy of the near-critical liquids under the electric field. Only recently, the new approach offering the common description of pretransitional effects in the isotropic liquid phase on approaching the nematic in LC and the orientationally disordered crystals (ODIC) in plastic crystalline (PC) materials was proposed [27, 28]. For the origins of Eq. 1 one also indicated the reasoning by Mistura [29] and Fisher [30], leading to the conclusion that for critical fluids, the isochoric heat capacity $C_V \propto d\varepsilon/dT \propto (T - T_C)^{-\alpha}$. The latter implemented to Eq. (1) yields:

$$\frac{d\varepsilon(T)}{dT} = a + A(1 - \alpha)|T - T^*|^{-\alpha} = a + A'|T - T^*|^{-\alpha} \qquad (2)$$

The coherent occurrence of Eqs. (1) and (2) were experimentally confirmed for $N, SmA \leftarrow I$ [18, 22], $SmA \rightarrow I$ [23, 24], and recently $SmA - N$ transitions [31]. It is also notable that the derivative-based analysis (Eq. 2) enable the local distortions-sensitive insight into the evolution described by Eq. (1). It also reduces the number of fitting parameter. The coherent analysis of both $\varepsilon(T)$ and $d\varepsilon(T)/dT$ dependences facilitates reaching optimal values of fitted parameters [31]. Notable that Eq. 2 contains fewer parameters than its hypothetical parallel for the heat capacity.

In wide-range temperature studies covering a few phases, selecting the proper reference frequency for determining the dielectric constant is essential. It requires broadband dielectric spectroscopy scans, presented as real and imaginary components of dielectric permittivity. Such results for undecylcyanobiphenyl (11CB) tested in the given report are shown in Figure 1.

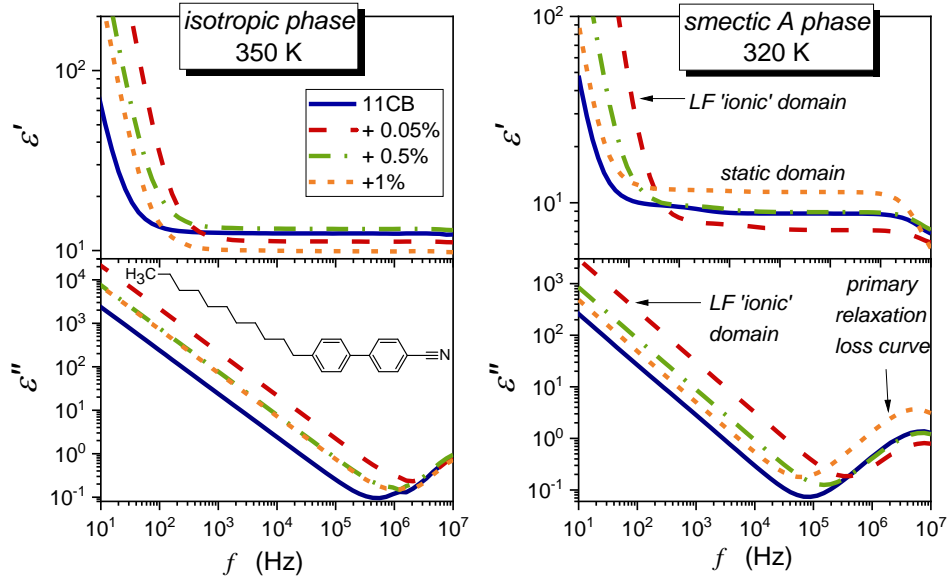

**Fig. 1** Examples of spectra detected by BDS - the real and imaginary parts of dielectric permittivity vs. frequency in 11CB and its BaTiO₃-based nanocolloids. Characteristic features of the spectrum and 11CB molecular structure are shown

The horizontal part of $\varepsilon'(f)$ is the static domain, related to the dielectric constant, i.e., $\varepsilon'(f) = \varepsilon = \varepsilon_S + \varepsilon_\infty$, where $\varepsilon_S$ is for the dipolar and $\varepsilon_\infty$ for electron molecular contributions. For high frequencies, $\varepsilon'(f)$ strongly decreases since the component $\varepsilon_S$ diminishes [16, 17]. The crossover frequency domain is associated with the strong manifestation of the primary loss curve $\varepsilon''(f)$. Its peak is related to the primary relaxation time $\tau = 1/2\pi f_{peak}$ associated with the molecular reaction on the electric field, and the height (maximum, $\varepsilon''_{peak}$) is the coupled energy absorption metric [16, 31]. Loss curve branches describe the distribution of relaxation time. All these properties are influenced by pretransitional fluctuations, at least in the isotropic liquid phase. The cognitive gap for the impact of pretransitional fluctuations on dielectric properties in the low-frequency (LF) domain still exists. This region is dominated by transport processes associated with ionic species, both $\varepsilon'(f)$, and $\varepsilon''(f)$ strongly increase on decreasing the frequency, as is visible in Fig.1. For discussing this behavior, two relations are explored in LC materials and their nanocomposites [32-40]:

$$\varepsilon'(f) = Af^{-3/2} + \varepsilon = \left(\frac{nq^2 D^{3/2}}{\pi^{3/2}\varepsilon_0 L k_B T}\right)f^{-3/2} + \varepsilon \tag{3}$$

$$\varepsilon''(f) = Bf^{-1} = \left(\frac{nq^2 D}{\pi \varepsilon_0 k_B T}\right)f^{-1} \tag{4}$$

where $n$ is the mobile ions concentration, $q$ denotes the charge of an ion, $D$ is the constant related to diffusion, $\varepsilon_0$ is the free space permittivity, $k_B$ is the Boltzmann constant, $L$ is the thickness of the dielectric cell (capacitor), and $T$ is the absolute temperature.

The validity of Eq. 4 is obvious when recalling the basic definition of DC electric conductivity, namely: $\sigma = \omega \varepsilon''(f) = 2\pi f \varepsilon''(f)$ [16, 17]. Distortions from Eq. 3 can yield insight into possible low-frequency relaxation or the Maxwell-Wagner polarization processes [16]. The authors have not found any reliable experimental test or explanation validating Eq.3, for the real part of dielectric permittivity in the LF domain.

To the best of the authors' knowledge, the only report focusing on the LF domain and the impact of pretransitional fluctuations considered the isotropic liquid phase of pentylcyanobiphenyl (5CB) [41]. It showed the following parameterizations for the temperature evolution [41]:

$$\varepsilon'(T, f < f_{static}) = [\varepsilon^* + a|T - T^*| + A|T - T^*|^{1-\alpha}]_{static} + [b|T - T^\wedge|]_{ionic} \qquad (5)$$

$$\Delta\varepsilon'(T, f) = \varepsilon'(T, f < f_{static}) - \varepsilon(T) = bT + bT^\wedge = bT + B \qquad (6)$$

where $b$ is the amplitude and $T^\wedge$ is the frequency-dependent singular temperature.

In the static domain, the last term in Eq. 5 diminishes. Notable that the linear behavior described by Eq. 6 terminates ~40 K above the clearing temperature, where fluctuations shrink to $2 - 3$ molecules [20, 21, 41].

This report develops the path indicated in ref. [41] for the isotropic liquid, also to the case of Smectic A, and solid crystal phase, and both in pure LC compound (11CB) as well as related nanocolloids. The results presented are supplemented by DC electric conductivity $\sigma(T)$ evolution insight, related to $\varepsilon''(f)$ behavior in the LF domain. Obtained BDS spectra enable examing the validity of Eq. (3).

## 2. Experimental

Studies were carried out in 4-undecyl-4'-cyanobiphenyl (11CB) with $Crystal \leftrightarrow 326.15K \leftrightarrow SmA \leftrightarrow 330.16K \leftrightarrow Nematic \leftrightarrow 330.65K \leftrightarrow Isotropic$ mesomorphism [1] The compound was synthesized and deeply cleaned by the LC team at the Military University of Technology, Warsaw, Poland. 11CB molecule is approximately rod-like, as shown in Fig.1, and has a permanent dipole moment parallel to the long molecular axis: $\boldsymbol{\mu}$ = 4.78 D [42]. Nonlinear dielectric effect (NDE) studies enable the reliable estimation of the phase transition temperature and discontinuity of the isotropic liquid to LC mesophase phase transition ($\Delta T$ = 5.6 K) via the linear regression fit to the temperature dependence of the reciprocal of NDE [43].

Paraelectric BaTiO$_3$ nanopowder (diameter $d$ = 50 nm) was purchased from US Research Nanomaterials, Inc. [44]. Mixtures of liquid crystal and nanoparticles were sonicated at a temperature higher than the isotropic to nematic phase transition for 4 hours to obtain homogeneous suspensions. Concentrations of nanoparticles are given in weight fraction percentage (wt%).

Samples were placed in a flat-parallel capacitor (diameter $2r$ = 20 mm) made of Invar with the distance between plates $d$ = 0.15 mm. The voltage of the measuring field $U$ = 1 V was applied. BDS studies were carried out using Novocontrol Alpha-A analyzer in the frequency range from 1 Hz to 10 MHz, with the Quatro Novocontrol unit for temperature control. Examples of dielectric permittivity spectra for the tested compound are shown in Figure 1. It also indicates significant features of such spectra. As the reference frequency for determining the dielectric constant $f$ = 116 $kHz$ was chosen: $\varepsilon = \varepsilon'(f = 116 kHz)$.

Studies of rod-like LC compounds are often carried out under conditions yielding insight into components of the dielectric constant - perpendicular ( $\varepsilon_\perp$ ) and parallel ( $\varepsilon_{//}$ ) to the short and long molecular axis of the molecule. It is reached via the use of the flat-parallel capacitor in which the tested sample is oriented by the (very) strong magnetic field. The alternative experimental path is a measurement in the capacitor with an adequately treated plate surface, supporting the required orientation of rod-like molecules and a tiny, micrometric gap between plates [1, 2, 4]. However, the orientation via an external magnetic field is possible only in the nematic phase. Thin-layer measurements, forcing the preferred direction, are possible in any phase, but the introduced constraints change generic features of the isotropic and smectic phases. Consequently, such studies are beyond the scope of the given report, which focuses on the LF properties of the 'native' phase, which is possible only using large ('bulk') capacitor gap and non-treated plates.

### 3. Results and discussion

Figures 2 shows temperature evolutions of dielectric constant in 11CB and its nanocolloids. The detailed behavior focused on the isotropic liquid phase is presented in Figure 3. The negligible impact of nanoparticles on the clearing temperature is notable, although a set of reports shows an apparent shift [14, 15]. However, the shift appears in nanocomposites with the third component – a molecular surface agent connected to the nanoparticles to avoid sedimentation. Consequently, several 'free' molecular dopants must also exist in such a three-component system. Notable that there is explicit experimental evidence that a molecular dopant in LC systems strongly changes the clearing temperature, also yielding the two-phase region between the isotropic liquid and LC mesophase [45, 46]. In this study, a surface agent is absent due to the focus on small 'colloidal' concentrations of NPs. Notable that a similar lack of a $T^C$ shift in nanocolloids composed solely from an LC compound and nanoparticles was reported earlier [23, 24, 31, 47, 48]. The lack of the third macromolecular component also prevented disturbances introduced into the BDS spectrum.

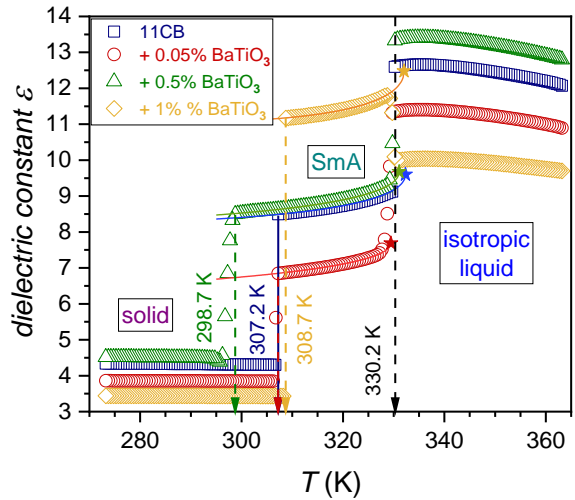

**Fig. 2** Temperature evolutions of dielectric constant in 11CB and 11CB+BaTiO₃ nanocolloids. Arrows indicate characteristic temperatures $T^C$ and $T_m$. The real part of the dielectric permittivity measured at $f = 116\ kHz$ was chosen as the dielectric constant ($\varepsilon'(T, f = 116\ kHz) = \varepsilon$). Curves portraying data are related to Eq. (1), with parameters given in Table 2. Stars show the hypothetical continuous phase transition point

However, the significant influence of nanoparticles on the SmA - Solid Crystal transition appears. It can mean that the disordering introduced by the nanoparticles can be substantial: it takes place in the structured SmA phase but it is absent in the disordered Isotropic Liquid. Notable that despite the paraelectric nature of added BaTiO$_3$ nanoparticles, they have a strong impact on the total value of the dielectric constant, mainly via the constant factor shift, related to $\varepsilon^*$ in Eq. (1), as it is shown explicitly below. This shift is significant even for a tiny concentration of NPs, $x = 0.05\%$. For the isotropic liquid and solid crystal phase, the addition of 0.5% of nanoparticles led to an increase in the dielectric constant value in comparison to the pure 11CB (about 17% in the isotropic phase), but when reaching $x = 1\%$, the substantial drop of $\varepsilon$ is visible.

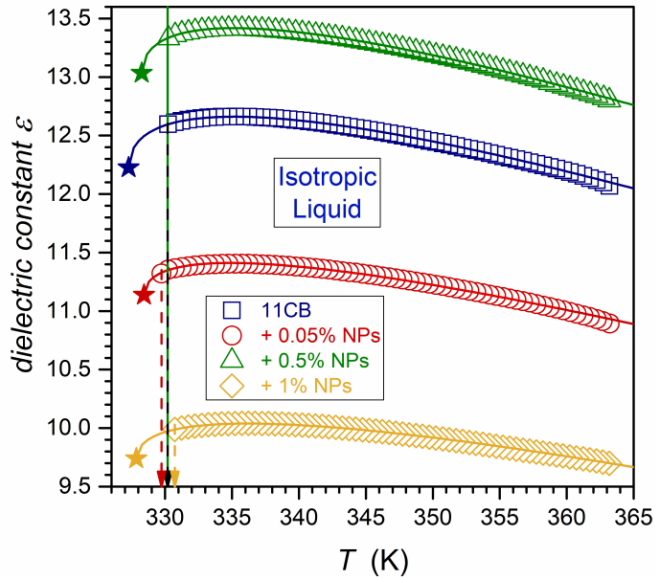

**Fig. 3** Temperature dependencies of dielectric constant in the isotropic liquid phase of 11CB and its nanocolloids with BaTiO$_3$. Curves portraying data are related to Eq. (1), with parameters given in Table 1. Stars show the hypothetical continuous phase transition points. Arrows indicate clearing temperatures for Isotropic Liquid – LC mesophase transitions

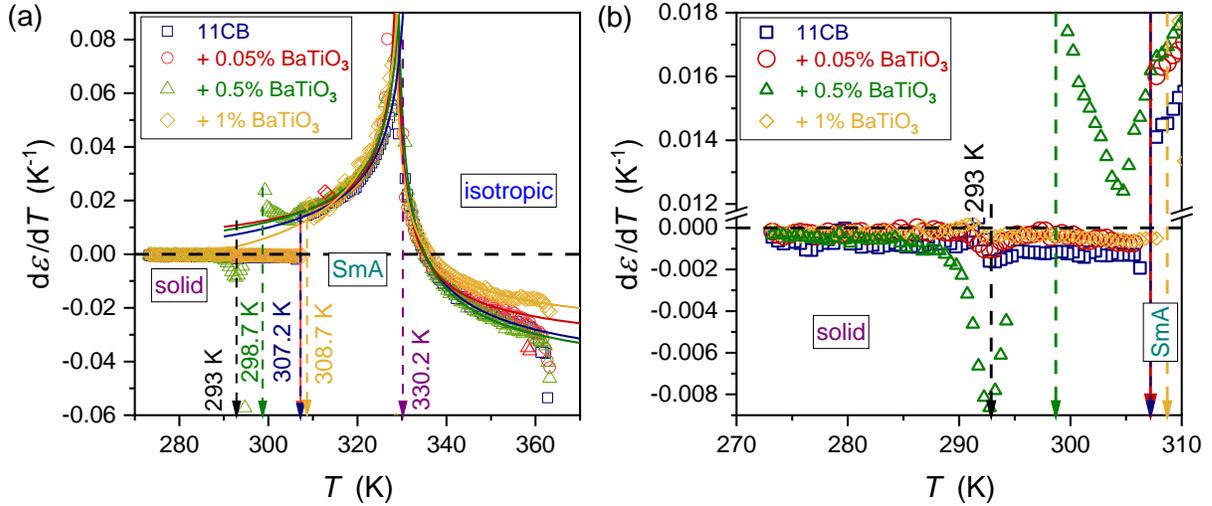

**Fig. 4** The derivative of experimental data for the temperature evolution of dielectric constant in 11CB and related BaTiO₃-based nanocolloids, given in (Fig. 2). The behavior in all tested phases is shown in Fig. 4a, and the focus on $Solid - SmA$ transition is presented in Fig. 4b.

In the SmA phase, the pattern is different. The addition of nanoparticles first decreases the dielectric constant by ~20% ($x = 0.05\%$), but for $x = 1\%$ dielectric constant value increases up to ~ 27% above that in pure 11CB.

Figure 4 presents the derivative of experimental data given in Fig. 2. The plot reveals the range of pretransitional effects associated with $N \longleftrightarrow I$, and $SmA \longleftrightarrow N$ phase transition: they cover the whole tested range of the isotropic liquid and smectic A phases, both in pure 11CB and nanocolloids. The nematic gap is too small (~0.5K) for a reliable test of pretransitional behavior, and it plays a role of an interesting 'disturbation' in the tests. The analysis of pretransitional effects was carried out by the simultaneous treatment of both $\varepsilon(T)$ and $d\varepsilon(T)/dT$ evolutions, given in Figs 2 and 4, respectively. Such an approach significantly increased the reliability of the nonlinear, multi-parameter fitting exploring Eqs. 1 and 2. Their results are shown graphically by solid curves in Figures 2-4, with parameters collected in Table 1 for $N \longleftarrow I$ transition and for $SmA \rightarrow N$ transition Table 2. For $N \longleftarrow I$ transition notable is the strong influence of NPs on the phase transition discontinuity $\Delta T^*$. For all tested systems, in the isotropic phase, the fit was optimal for the exponent value $\alpha = 0.5$, which is the hallmark feature for two 'classic' approaches: mean-field and tricritical point (TCP) related. The latter is experimentally supported by the results of dielectric constant studies focused on the order parameter in 5CB and 8OCB [21, 29, 49] also containing the derivative analysis. They yielded the order parameter exponent TCP value $\beta = 1/4$ [21, 49], Notable that for the mean-field model described systems $\beta = 1/2$. The value of the exponent $\alpha = 1/2$ for the isotropic liquid phase is well established experimentally by dielectric constant, heat capacity and related studies [5-10, 18-24, 29-31].

The evidence for $SmA \rightarrow N$ is more complex. Already de Gennes, in his classic monograph [11], indicated the parallel between phase transitions in superconductors and $N - SmA$ in LC materials. For the *Physics of Critical Phenomena,* it means that both systems, different at the microscopic level, belong to the same universality class XY(2D) Heisenberg [4]. It is related to the specific heat critical exponent $\alpha \approx -0.007$. However, experimental results of heat capacity measurements for N-SmA transition are puzzling. The obtained values of the exponent range from $\alpha \approx -0.007$ to $\alpha \approx 0.5$ [4, 50-52]. It is explained via the impact of the Fisher renormalization [53, 54], associated with the presence of an additional component, or the empirical correlation to the 'temperature width' of the nematic phase, located between the isotropic liquid and SmA phase [50]. In the given report, the value $\alpha \approx 0.5$ and the slightly discontinuous character of the $SmA \rightarrow N$ transition offers a fair portrayal of experimental data (Table 2) in all tested samples. However, nanoparticles notably influence the discontinuity $\Delta T^*$. The authors would like to draw attention to the recent evidence showing a split of the nematic phase in 8OCB into two domains ($N_I$, $N_{SmA}$), associated with neighboring of $I - N$ and $N - SmA$ transitions, respectively [31]. Notable is the 'width' of the nematic phase in 8OCB: $\Delta T_N \approx 12K$ and the obtained exponent $\alpha = 0.11 - 0.2$ [31]. For 11CB the nematic phase width is tiny: $\Delta T_N \approx 0.5K$. Consequently, one can consider rather phase transitions: $N_I - SmA$ for 11CB, and $N_{SmA} - SmA$ for 8OCB. It can offer the additional comment regarding the origins of critical exponent values for $N - SmA$ transition in different systems.

**Table 1** Results of portraying dielectric constant change in the isotropic phase of 11CB and its nanocolloids with $BaTiO_3$ via Eq. (1) and Eq. (2). These results are shown graphically in Fig. 3 and Fig.4.

| | *11CB* | *+ 0.05 wt%* | *+ 0.5 wt%* | *+ 1 wt%* |
|---|---|---|---|---|
| $\varepsilon^*$ | 12.2 | 11.1 | 13.0 | 9.7 |
| $T^C$ (K) | 330.2 | 329.7 | 330.2 | 330.7 |
| $T^*$ (K) | 327.3 | 328.4 | 328.3 | 327.9 |
| $\Delta T$ (K) | 3.0 | 1.3 | 2.0 | 2.9 |
| $a$ | -0.055 | -0.043 | -0.056 | -0.036 |
| $A$ | 0.31 | 0.22 | 0.29 | 0.21 |
| $\alpha$ (fixed) | 0.5 | 0.5 | 0.5 | 0.5 |

**Table 2** Parameters for $N$ $SmA$ pretransitional effect portrayed via the coherent analysis using Eqs. (1) and (2). These results are shown graphically in Fig. 2 and Fig.4

| | *11CB* | *+ 0.05 wt%* | *+ 0.50 wt%* | *+ 1 wt%* |
|---|---|---|---|---|
| $\varepsilon^*$ | 9.60 | 7.69 | 9.67 | 12.48 |
| $T^C$ (K) | 330.2 | 329.7 | 330.2 | 330.7 |
| $T^C - 0.5 = T_{N\text{-}SmA}$ (K) | 329.7 | 329.2 | 329.7 | 330.2 |
| $T^*$ (K) | 332.3 | 329.5 | 331.1 | 332.0 |
| $\Delta T$ | 2.6 | 0.3 | 1.4 | 1.8 |
| $a$ | 0.017 | 0.006$_4$ | 0.013 | 0.030 |
| $A$ | -0.31 | -0.21 | -0.28 | -0.42 |
| $\alpha$ (fixed) | 0.5 | 0.5 | 0.5 | 0.5 |

Figure 4b focuses on the solid phase, which has been hardly discussed so far. Generally, for the melting/freezing transition, no pretransitional effects are expected [55]. Such behavior is also evidenced for LC-based materials discussed in the given report, except $x = 0.5\%$ nanocolloid, where a notable pretransitional effect appears in SmA (fluid) mesophase. Moreover, for this nanocolloid, the additional Solid-Solid transition also emerges. In the authors' opinion, the Solid phase is probably associated with plastic crystalline phases. Such a statement can support the mean value of the dielectric constant $\varepsilon \approx 4.5$ in this domain: for a solid crystal with frozen translational and orientational arrangements, one can expect $2 < \varepsilon < 2.5$. The latter is also evidenced in Fig. 2, but solely for $x = 1\%$ nanocolloid.

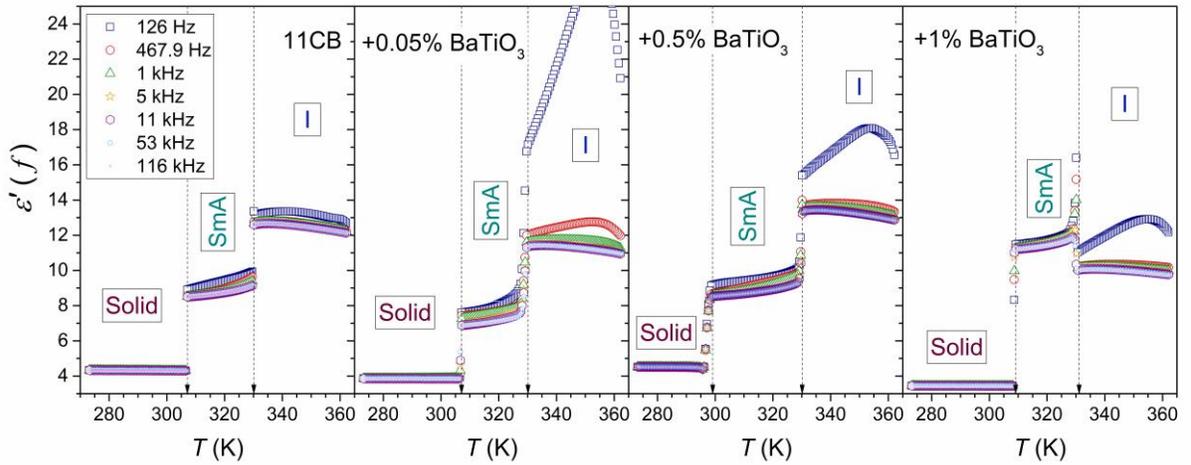

**Fig. 5** The real part of dielectric permittivity for a set of frequencies in the low-frequency domain in 11CB and its nanocolloids, from the static domain related to $f = 116kHz$ to $f = 126Hz$ in the LF domain.

Premelting effects on the solid side of the melting temperature $T_m$ are relatively often observed [55, 56]. They are not linked to the pretransitional phenomena, but the quasi-liquid channels between solid grains forming near $T_m$ [57]. The distortions-sensitive analysis in Fig. 4b reveals such behavior in systems discussed in this report and the possibility of its strengthening by the added nanoparticles.

Figure 5 presents the behavior of the real part of dielectric permittivity for a set of frequencies, from the static to the low-frequency (LF) domain. Nanoparticles strongly influence $\varepsilon'(f)$ value, when decreasing the monitoring frequency. This is particularly visible in the isotropic liquid phase. This increase in $\varepsilon'(f)$ value is associated with the apparent disappearance of the above-mentioned pretransitional effect.

Notwithstanding, the impact of pretransitional fluctuations becomes explicitly visible if data from Fig. 5 are considered in frames of the 'difference analysis', introduced by Eq. 6, and yielding the insight limited to the ionic-related contribution. Such results are presented in Fig. 6 for all tested phases of 11CB and its nanocolloids. Figures 7 and 8 show the behavior focused on the isotropic liquid and solid phases. For the isotropic phase, the linear behavior, suggested in Eqs. 5 and 6, is visible. Such behavior is strongly enhanced when adding nanoparticles. Even a tiny addition of NPs (0.05%) can significantly increase the ionic contribution to dielectric permittivity in comparison to pure 11CB, and this impact diminishes when increasing NPs concentration.

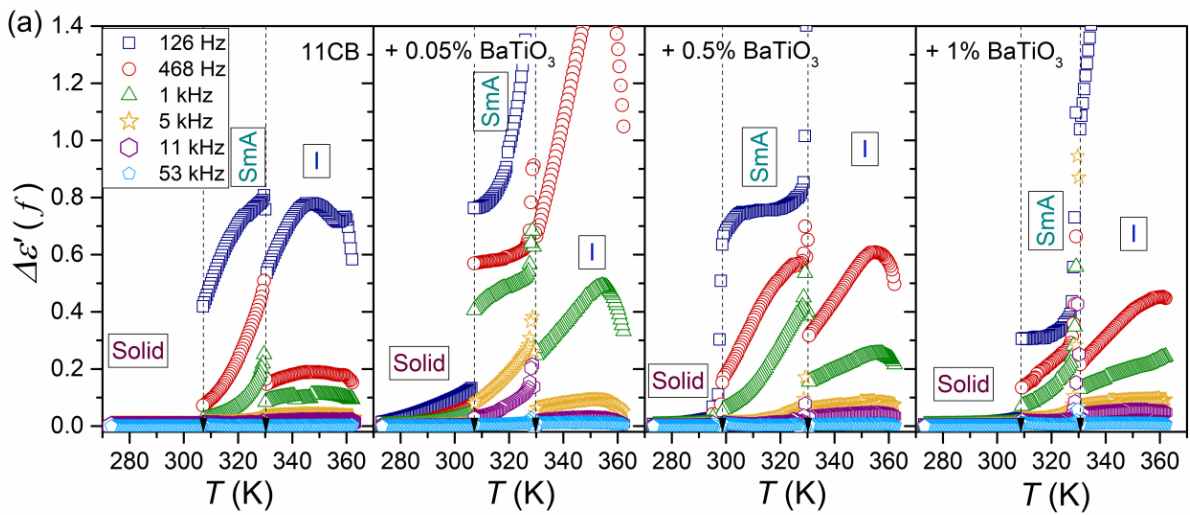

**Fig. 6** Temperature changes of the residual-ions contribution to the real part of dielectric permittivity emerging on decreasing frequency in the LF region, below the static domain; $\Delta\varepsilon'(f) = \varepsilon'(f) - \varepsilon$, in all tested phases of 11CB and its nanocolloids

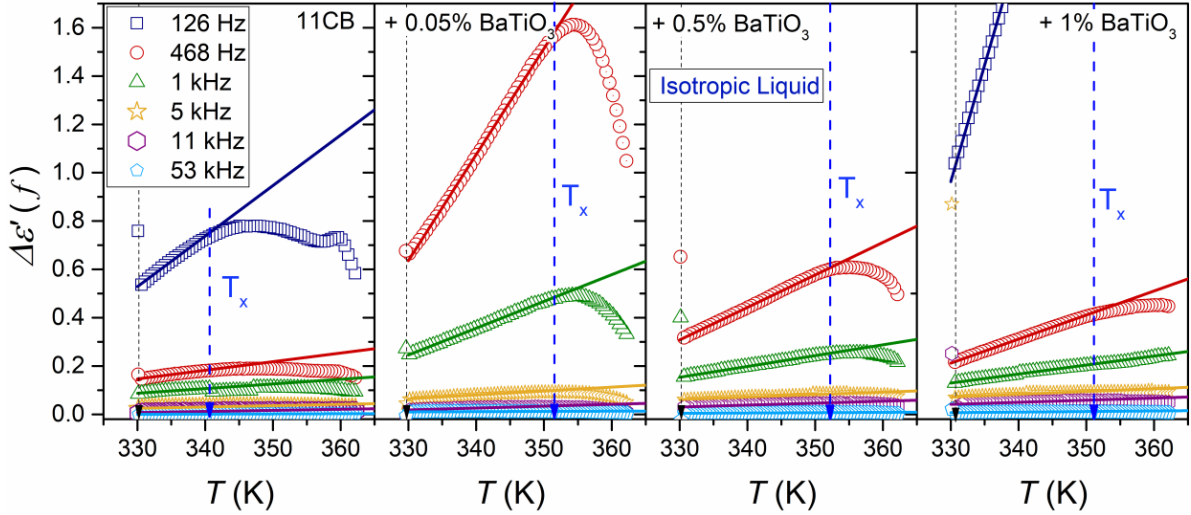

**Fig. 7** The focus on the isotropic liquid phase of 11CB and its nanocolloids: temperature changes of the residual-ions contribution to the real part of dielectric permittivity emerging on decreasing frequency in the LF region, below the static domain; $\Delta\varepsilon'(f) = \varepsilon'(f) - \varepsilon$

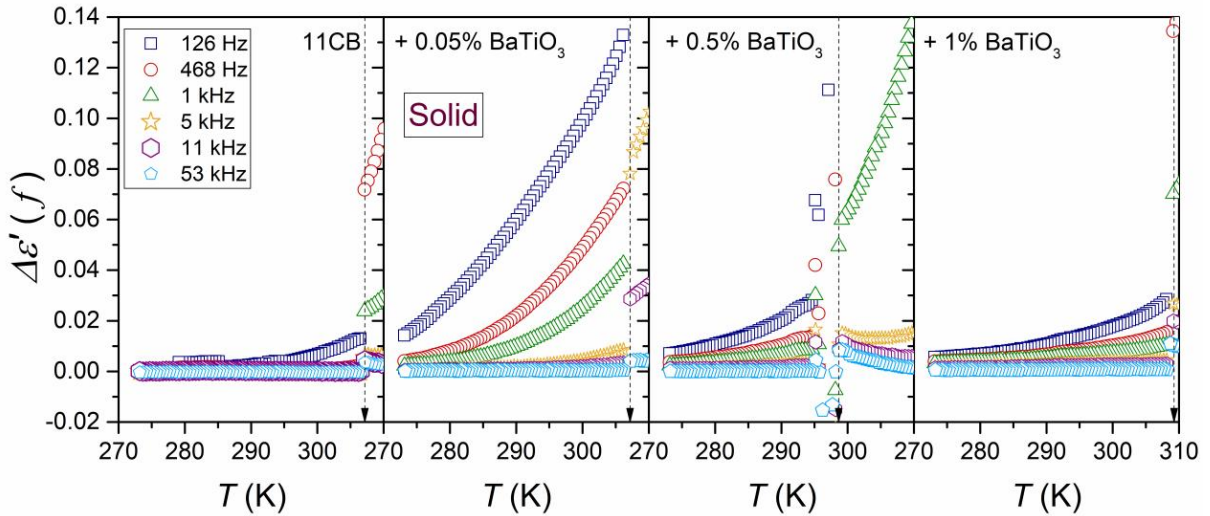

**Fig. 8** The focus on the solid phase of 11CB and its nanocolloids.: temperature changes of the residual-ions contribution to the real part of dielectric permittivity emerging on decreasing frequency in the LF region, below the static domain; $\Delta\varepsilon'(f) = \varepsilon'(f) - \varepsilon$

In the SmA phase, notable are (very) strong pretransitional changes of $\Delta\varepsilon'(f,T)$ for $SmA \to N$ transition. The discussed way of analysis also leads to the emergence of pretransitional effects for $Solid\ (crystal) \leftarrow (fluid)\ SmA$ transition, as well as $Crystal \to SmA$. One should recall that $\Delta\varepsilon'(f,T)$ is related solely to the ionic contribution of the electric polarizability, whereas permanent dipole moments contribute mainly to the dielectric constant.

When discussing dielectric properties in the low-frequency domain, one cannot omit DC electric conductivity, being a 'dynamic equivalent' of the dielectric constant, related to $\varepsilon''(f)$, as follows [16, 17]:

$$\sigma_{DC} = \sigma = \omega\varepsilon''(f) = 2\pi f\,\varepsilon''(f) \tag{7}$$

It means that a straight line with a slope of –1 should appear on the log-log scale in the LF part of $\varepsilon''(f)$ spectrum, as in Fig. 1.

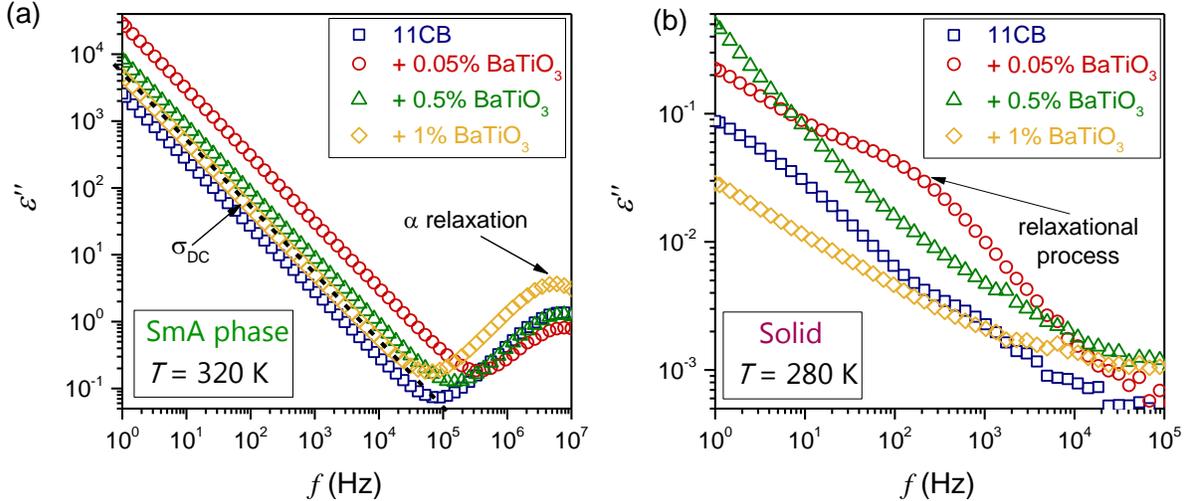

**Fig. 9** The imaginary part of dielectric permittivity spectra focused on the low-frequency domain for SmA phase (9a) and the solid phase (9b) in 11CCB and its nanocolloids with BaTiO$_3$.

Figures 9(a, b) present such behavior for the SmA mesophase and solid phase in 11CB. For the SmA phase, Eq. 7 is perfectly fulfilled, but for the solid phase, it is distorted by LF relaxational processes and the Maxwell-Wegner polarization effect. In such a case, Eq. 7 cannot be used for estimating DC electric conductivity. The ultimate test for the reliable determining DC electric conductivity constitutes the transformation for the complex dielectric permittivity representation, as shown in Figure 10. The horizontal part in the plots, for SmA and Isotropic liquid phase, is for DC electric conductivity.

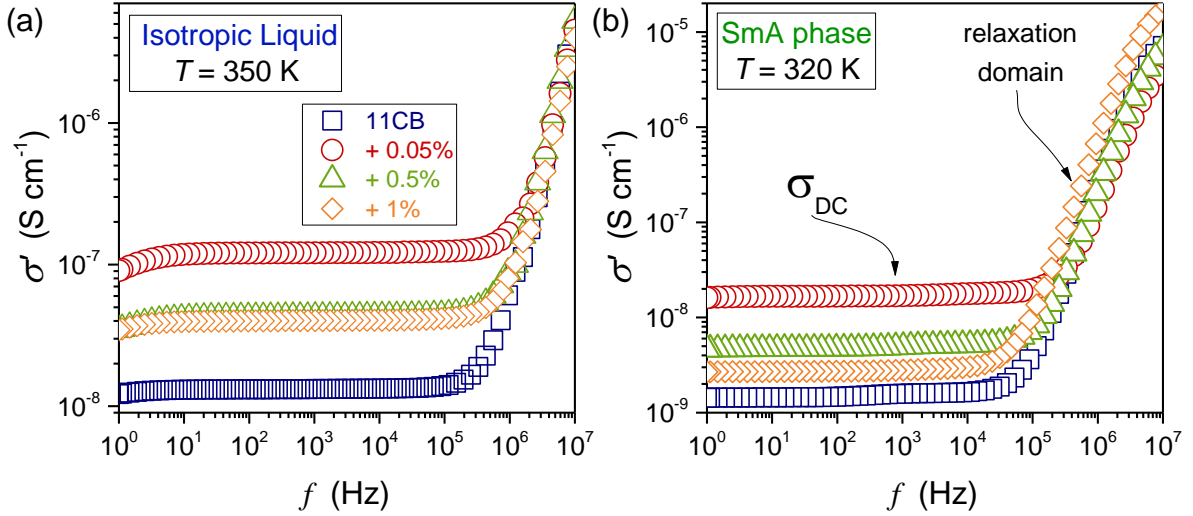

**Fig. 10** The real component of the complex electric conductivity obtained from data shown in Fig. for the Sm A phase and the isotropic liquid phase. Results were obtained via Eq. 9, for 11CB and its nanocolloids. The 'stationary' (frequency-independent) and relaxation domains are indicated.

The primary relaxation time and DC electric conductivity are linked via the Debye-Stokes-Einstein (DSE) relation [58]:

$$\tau(T)\sigma(T) = C \quad \Rightarrow \quad \frac{\tau(T)}{C\sigma^{-1}(T)} = 1 \quad \Rightarrow \quad \frac{\tau(T)}{\tau_\sigma(T)} = 1 \qquad (8)$$

The right-hand part of Eq. 8 is for the basic DSE equation. In complex systems, particularly near the glass transition, the exponent showing the translational-orientational (T&O) decoupling can appear, associated with the exponent S coupled to the orientational relaxation time. The right-hand part of Eq. 8 shows that the reciprocal of DC electric conductivity is the metric of the translational ($\sigma$- related) relaxation time. Consequently, the following general evolution of discussed properties can be expected [16]:

$$\sigma_{DC}^{-1}(T), \tau(T), \tau_\sigma(T) \propto exp\left(\frac{E_a(T)}{RT}\right) \qquad (9)$$

where $E_a(T)$ is the apparent (temperature-dependent) activation energy, and $R$ stands for the gas constant.

Eq. 9 is simplified to the basic Arrhenius pattern if $E_a(T) = E_a = const$, in the given temperature domain. In such a case, one obtained the linear domain for the so-called Arrhenius plot [16]. For instance:

$$log_{10}\sigma^{-1}(T) = \frac{ln\sigma_0^{-1}}{ln10} + \frac{E_a}{ln10RT} = A + B \times \frac{1}{T} \qquad (10)$$

where $\sigma_0^{-1}$ is the prefactor in Eq. 9; constant parameters: $A = ln\sigma_0^{-1}/ln10$ and $B = 1/Rln10$.

If the T&O decoupling takes place, in Eq. 9: $E_a^\sigma(T) = SE_a(T)$ appears [59]. Figure 11 shows the Arrhenius plot for the reciprocal of DC electric conductivity in the isotropic liquid and SmA phase of 11CB and its BaTiO$_3$-based nanocolloids. The linear behavior in the SmA phase indicates the basic Arrhenius pattern, with the constant activation energy and a minor pretranstional distortion near the N-SmA transition

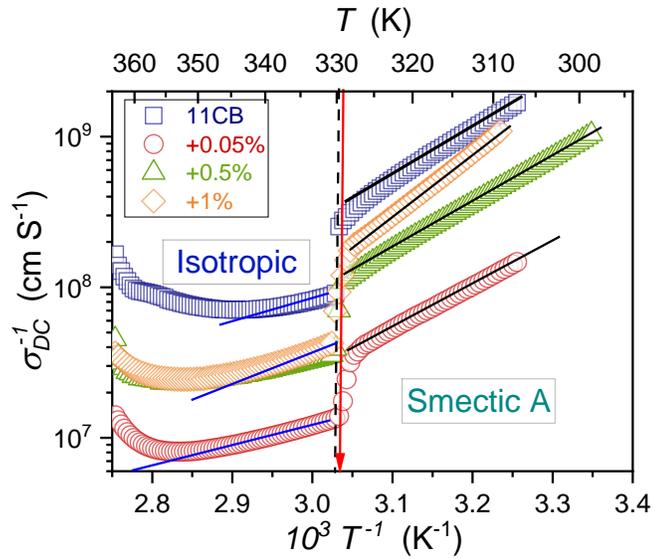

**Fig. 11** Evolutions of reciprocals of DC electric conductivity in the isotropic and SmA phase of 11CB and its nanocolloids.

For the isotropic liquid, such behavior is visible only for $x = 0.05\%$ nanocolloid. For other tested nanocolloids and pure 11CB, the Super-Arrhenius behavior with apparent (changeable) activation energy takes place. It can be linked to the joined impact of pretransitional, pre-LC, fluctuations, and nanoparticles. Following Eqs. 9 and 10, one should expect the systematic rise of $\sigma^{-1}(T)$, $\tau_\sigma(T)$, and $\tau(T)$ on cooling. Indeed, such behavior is evidenced in Fig. 11, except for the high-temperatures limit in the isotropic liquid. The authors do not have an explanation for this phenomenon.

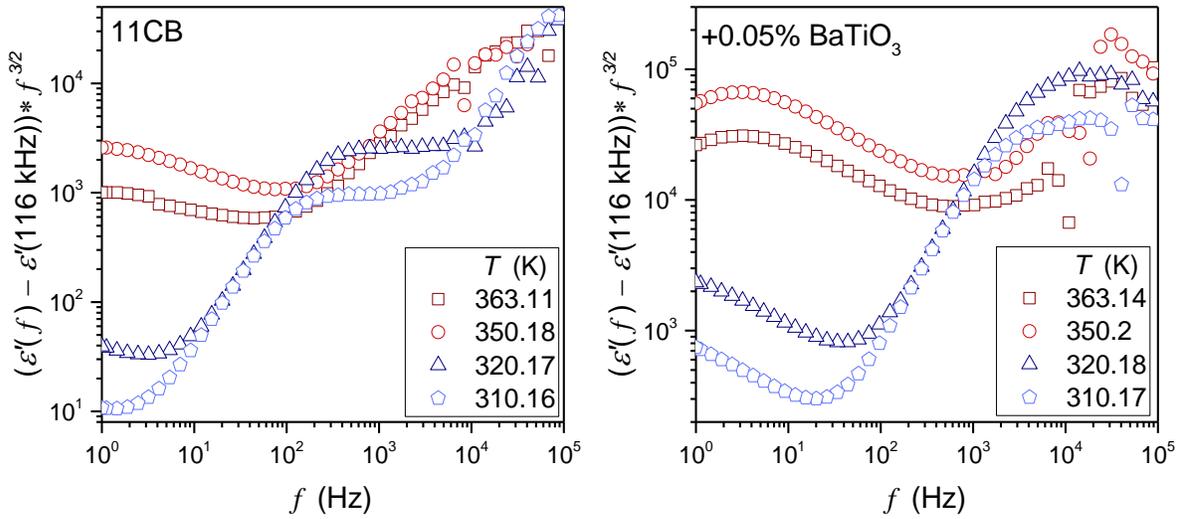

**Fig. 12** The test of Eq. 12, suggesting the universal frequency-related behavior of the real part of dielectric permittivity in the ionic-dominated LF domain (Eq. 3): results for 11CB and its nanocolloid in the isotropic liquid, SmA, and Solid phases. The lack of horizontal behavior clearly shows the lack of the universality suggested by Eq. 3.

Finally, we would like to address Eqs. (2) and (3), which are often recalled when discussing dielectric spectra in the LF domain [32-40]. The above discussion explicitly validated Eq. 3, showing its link to DC electric conductivity and indicating validity limits. For Eq. 4 the authors have not found experimental validating results. Notwithstanding, one can propose the simple scaling plot enabling the ultimate validation, based on Eq. 4:

$$\Delta\varepsilon'(f)f^{3/2} = A = const \tag{11}$$

where $\Delta\varepsilon'(f) = \varepsilon'(f) - \varepsilon$.

Figure 12 presents the results of the scaling analysis of Eq. 3 based on Eq. 11. No horizontal line, which should appear if Eq. 3 is valid, is visible. Hence, the test of Eq. 3 is negative.

## 3. Conclusions

This report shows the analysis of the temperature behavior of the real part of dielectric permittivity in the static (dielectric constant) and low-frequency domains in bulk samples of 11CB and its BaTiO$_3$-based nanocolloids, extending from the isotropic liquid, Smectic A to the solid phase. In each phase, the dominated impact of pretransitional fluctuations, significantly moderated by nanoparticles, was detected. The authors propose to split focus on the dielectric constant $\varepsilon(\boldsymbol{T})$ evolution, yielding mainly response from permanent dipole moment and their arrangement, and in the LF domain $\Delta\varepsilon'(\boldsymbol{f}) = \varepsilon'(\boldsymbol{f}) - \boldsymbol{\varepsilon}$ which is associated solely with ionic-related polarization mechanisms. Notable is also the distortions sensitive analysis, offering insight into local distortion within the tested temperature dependence. All of these led to new experimental evidence for I-N, N-SmA and SmA-Solid transitions in tested systems: for the extent of pretransitional effects (i.e., fluctuations-dominated domains), critical exponent, and phase transitions discontinuities. Notable is the evidence of pretransitional effects for the SmA – Solid Crystal transition, particularly for $\Delta\varepsilon'(\boldsymbol{f},\boldsymbol{T})$ evolutions. This issue is worth stressing since the transition from the LC mesophase to the solid phase still constitutes a cognitive gap, in the authors' opinion. Finally, we would like to stress the validating discussion of Eqs. 3 and 4, which are often recalled as the reference in discussing complex dielectric permittivity in the low-frequency domain.

**Conflicts of interest**

There are no conflicts of interest for the authors of this report

## Author Contributions

JŁ was responsible for measurements, data analysis, and paper preparation, ADR supported data analysis and paper preparation, SJR supported paper preparation. KC carried out the synthesis of LC samples, purification, and the basic mesomorphism test.

## Acknowledgments

This work was supported by the National Science Center (NCN Poland), grant numbers: 2017/25/B/ST3/02458, the head: S. J. Rzoska; and 2016/21/B/ST3/02203, headed by A. Drozd-Rzoska. The paper is associated with the *International Seminar on Soft Matter & Food – Physico-Chemical Models & Socio-Economic Parallels*, *1st Polish-Slovenian Edition*, Celestynów, Poland, 22–23 Nov. 2021; directors: Dr. hab. Aleksandra Drozd-Rzoska (Institute of High Pressure Physics PAS, Warsaw, Poland) and Prof. Samo Kralj (Univ. Maribor, Maribor, Slovenia).